\documentstyle[12pt, aasms4]{article}
\begin{document}

\vspace*{2.0in}
\title{Extended dust emission  and atomic hydrogen, \\ a reservoir of diffuse
H$_2$ in NGC 1068}

\author{Padeli. P. Papadopoulos}
\affil{Sterrewacht Leiden, P. O. Box 9513, 2300 RA Leiden, The Netherlands}
\and
\author{E. R. Seaquist}
\affil{Department of Astronomy, University of Toronto, 60 St. George st.
       Toronto,\\ ON M5S$-$3H8, Canada.}

\begin{abstract}

We report  on sensitive sub-mm  imaging observations of  the prototype
Seyfert~2/starburst galaxy  NGC 1068  at 850 $\mu  $m and 450  $\mu $m
using  the Submillimetre  Common-User Bolometer  Array (SCUBA)  on the
James Clerk Maxwell Telescope (JCMT).   We find clear evidence of dust
emission associated with the extended HI component which together with
the very faint $ ^{12}$CO  J=1--0 emission give a gas-to-dust ratio of
$\rm M_{\rm  gas}/M_{\rm dust} \sim 70-150$.  This  contrasts with the
larger ratio $\rm M_{\rm gas}/M_{\rm dust}\sim 330$ estimated within a
galactocentric radius  of $r\leq  1.36$ kpc, where  the gas  is mostly
molecular and starburst activity  occurs.  The large gas-to-dust ratio
found  for  the  starburst   region  is  attributed  to  a  systematic
overestimate of the molecular  gas mass in starburst environments when
the luminosity of  the $ ^{12}$CO J=1--0 line  and a standard galactic
conversion factor is used.  On the other hand sub-mm imaging proves to
be a more powerful tool than conventional CO imaging for revealing the
properties  of the  diffuse $\rm  H_2$  that coexists  with HI.   This
molecular  gas   phase  is   characterized  by  low   densities  ($\rm
n(H_2)<10^3$  cm$  ^{-3}$),  very  faint emission  from  sub-thermally
excited   CO,  and   contains   more  mass   than   HI,  namely   $\rm
M(H_2)/M(HI)\sim 5$.

\end{abstract}


\section{Introduction}

The standard way  to estimate molecular gas mass  in galaxies uses the
 luminosity of the $ ^{12}$CO J=1--0 line and converts it to molecular
 gas mass  by using the so-called standard  galactic conversion factor
 $\rm  X_{\rm co}$  (e.g.  \cite{YoSco82};  \cite{Blo85}; \cite{Di86};
 \cite{YoSco91}).  The  dependence of $\rm X_{\rm co}$  on the ambient
 conditions of  the molecular gas has been  extensively explored.  The
 main factors are:  a) metalicity and the intensity  of the ambient UV
 radiation   field  (e.g.   \cite{Is88};   \cite{Wil95};  \cite{Ar96};
 \cite{Is97}), b) the density, temperature, and kinematic state of the
 average  molecular cloud  and,  c) effects  of particular  geometries
 (e.g. see \cite{Br96}; \cite{Sak96} for a recent exposition).  In the
 environments of  extreme starbursts it has  been clearly demonstrated
 that the  standard method overestimates $\rm M(H_2)$  since it yields
 masses   comparable  or   larger  than   the  dynamical   ones  (e.g.
 \cite{Do93};   \cite{Br96};   \cite{So97};   \cite{Do98}).   In   low
 metallicity environments with high  UV intensities the larger rate of
 CO  dissociation  with respect  to  $\rm  H_2$  produces a  lower  CO
 luminosity per $\rm H_2$ and thus underestimates $\rm M(H_2)$.

In principle the aforementioned effects can be taken into account by a
combination  of observational  and  theoretical work  studying a  wide
range of  galactic environments,  which can then  provide ``adjusted''
$\rm  X_{\rm  co}$ factors  to  be used  for  the  appropriate set  of
conditions.  However little can be done if the $\rm H_2$ is so diffuse
and/or cold that  the $ ^{12}$CO J=1--0 line is very  faint or not not
luminous at all.  Hints of such physical conditions have been recently
reported for  the outer parts  of Milky Way  (Usuda et al.   1998) and
cold dense clouds have been  reported for M 31 (\cite{Al98}).  In such
cases the advantage  of easily detecting and mapping  a bright CO line
and thus  the distribution of $\rm  H_2$ is lost. Here  we report deep
sub-mm  imaging of the  prototype Sy2/starburst  galaxy NGC  1068 that
reveals extended emission from  dust well beyond the CO-bright regions
and associated with the HI.

\section{Observations}

The observations  were made on  two nights in  1997 August 4  and 1998
January 26 with the Sub-mm  Common User Bolometer Array (SCUBA) at the
15  m  James  Clerk  Maxwell  Telescope  (JCMT)\footnote{The  JCMT  is
operated by  the Joint Astronomy Center  in Hilo, Hawaii  on behalf of
the  parent organizations PPARC  in the  United Kingdom,  the National
Research Council  of Canada and  the The Netherlands  Organization for
Scientific  Research.}.  SCUBA  is a  dual camera  system  cooled well
below 1 K that allows sky-background limited simultaneous observations
with two  arrays.  The short-wavelength  array contains 91  pixels and
the long-wavelength  array 37 pixels.  Both  arrays have approximately
the same field of view on  the sky $\sim 2.3'$. For a full description
of the instrument see \cite{Ho98}.

We performed  dual wavelength imaging at  450 $\mu $m and  850 $\mu $m
using the 64-point jiggle mapping mode that allows Nyquist sampling of
the  field of view  (\cite{Ho98}). We  employed the  recommended rapid
beam switching at a  frequency of 8 Hz and a beam  throw of $120''$ in
azimuth.   The pointing  and  focus of  the  telescope were  monitored
frequently  using Mars,  Jupiter and  CRL  618, with  an expected  rms
pointing error of $\leq 3''$.  All the NGC~1068 maps were bracketed by
skydips used to correct  for atmospheric extinction. Typical opacities
at 850 $\mu  $m were $\tau \sim 0.14$  for our 1998 run  and $\tau \sim
0.6$  for our 1997  run.  Frequent  photometric measurements  and beam
maps  of  CRL~618 and  Mars  allowed  close  monitoring of  the  gains
(Jy/beam Volt$  ^{-1}$).  These  can vary significantly  especially at
450 $\mu $m  if the dish  has not thermally  relaxed and can  even be
elevation dependent (Sandell~1998).

  The  individual jiggle  maps are  consistent  in terms  of peak  and
integrated   intensities  and   were   co-added  after   flatfielding,
correcting   for   atmospheric  extinction,   and   editing  out   bad
bolometers/integrations  using  the  standard reduction  package  SURF
(\cite{Jen98a}).   Special care  was taken  to remove  sky-noise using
bolometers  at the  edge of  the field  of view  that ``look''  at sky
emission  only  (\cite{Jen98b}).  As  a  result  the  final maps  have
exceptional  sensitivity, close  to the  one expected  from  the total
integration time and the NEFDs at the sky conditions of our runs.  Flux
calibration of the final maps  in mJy/beam was obtained from beam maps
of CRL~618 obtained with  the same chop  throw as our NGC  1068 maps.
The corresponding  beam profiles were  obtained from beam maps  of CRL
618 and Mars.  The calibration information is summarized in Table 1.

\vspace*{1.0cm}

\centerline{EDITOR: PLACE TABLE 1 HERE}
\vspace*{0.3cm}

\section{Extended sub-mm emission: CO, HI and dust}

The clear association  of the sub-mm emission from dust  and HI gas is
demonstrated in  Figure 1 where the 850  $\mu $m and 450  $\mu $m maps
are  overlayed with  an  HI map  (Brinks  et al.   1994)  at a  common
resolution of $\sim 15''$.  Figure 2 shows the bright central 450 $\mu
$m emission and the associated $ ^{12}$CO J=1--0 integrated brightness
(\cite{Hel95}) that lie within the  central HI ``hole'' seen in Figure
1.  This wealth of  imaging data for the ISM in  NGC~1068 and the wide
range  of conditions  present  due  to the  existence  of the  central
starburst (Telesco et  al.  1984) make it an  ideal testing ``ground''
of our standard  ideas about the distribution and  mass of the various
ISM phases on large scales.  We adopt a distance to NGC~1068 of 14~Mpc
($\rm H_{\circ } = 75 \ km\ s^{-1}\ Mpc^{-1}$) where $1''$ = 68 pc.

A remarkable property  of the sub-mm emission is  that it extends over
 the   entire   bright   HI   emission  where   the   estimated   $\rm
 R=S_{450}/S_{850}$ ratio is lower  than in the CO-bright region where
 the starburst occurs. This ratio  and its variations across the image
 can  be a  good temperature  indicator as  long as  the  average dust
 temperature is  $\rm T\leq 30$  K. Indeed, assuming a  single average
 dust component, this ratio can be expressed as

\begin{equation}
\frac{\rm S_{450}}{\rm S_{850}} = 1.88^{\beta + 3}\ \left( \frac{\rm e^{\rm 16.8/T}-1}{\rm e^{\rm 31.8/T}-1}\right),
\end{equation}

\noindent
where $\beta $  is the emissivity law. It can be  easily seen that the
ratio tends to the R-J limit  for dust temperatures of $\rm T\geq 30$K
and is no longer temperature-sensitive.  Nevertheless, for a range of
$\rm T=10-30$ K, it varies by  a factor of $\sim 2$.  This temperature
range is  particularly interesting  since it ``marks''  the transition
between cool and warm dust, the latter dominating the IRAS 100 $\mu $m
and 60 $\mu $m bands.

After carefully correcting the integrated flux densities for the error
beam  contribution   (Sandell  1998)  using   the  correction  factors
tabulated in  Table 1, we obtain  $\rm R(r\leq 20'')  = 7.10\pm 1.77$.
For $\beta  = 2$ this agrees  well with the range  of gas temperatures
$\rm T\sim 20-30$ K found  from studying the CO excitation within this
region  (\cite{Pap98}) and  the  minimum temperature  of  $\sim 20$  K
implied from the observed $ ^{12}$CO J=1--0 brightness temperatures in
high  resolution maps  (e.g.  \cite{Pla91}).  For  the sub-mm  emission
associated with  the HI  distribution we find  $\rm R(20''\leq  r \leq
60'') = 4.70\pm  1.17$.  This ratio suggests colder  gas, in the range
of $\rm T\sim 10-15$ K, but rules out any M 31-type clouds (Loinard \&
Allen 1998) where $\rm T_{\rm kin}<10$ K.  Here we must emphasize that
the main uncertainty in the estimate of R is due to systematic factors
($\sim 25\%$)  that stay unchanged  across the maps.  The  thermal rms
uncertainty is  $\leq 5\%$ and thus  the observed change  of R between
the CO-bright and the HI-bright  regions is much more significant than
the total quoted uncertainties imply.

\section{The gas-to-dust ratio}

The  spatial resolution  of our  sub-mm maps  and the  HI,  $ ^{12}$CO
J=1--0 maps  from the literature  permits us to estimate  the gas/dust
ratio  as  a function  of  position  in  this galaxy.   Combining  the
standard  expressions for  the HI,  H$_2$ and  dust mass  the gas/dust
ratio in astrophysical units is expressed as follows:

\begin{equation}
\frac{\rm M_{\rm g}}{\rm M_{\rm d}}\approx 1.25\times 10^3
 \left(\frac{\nu }{\nu _{\circ }}\right)^{\beta + 3} 
\left(\rm e^{\rm h\nu/k T}-1\right)^{-1}
\left[\frac{\rm S_{\rm HI} + 10^{-2} X_{\rm co}
 S_{\rm CO}}{\rm S_{\nu }}\right],
\end{equation}

\noindent
where   $\rm   S_{\rm   HI}$    and   $\rm   S_{\rm   CO}$   are   the
velocity-integrated flux densities of  the HI hyperfine transition and
the  $ ^{12}$CO  J=1--0 line  respectively in  Jy km  s$  ^{-1}$, $\rm
X_{\rm co}$  is the  standard galactic conversion  factor in  units of
$\rm M_{\odot }\ (K\ km \ s^{-1}  \ pc^2)^{-1}$ and $\rm  S_{\nu }$ is
the sub-mm  flux density at  frequency $\nu $  in~Jy.  We  adopted the
emissivity law given by Hildebrand  (1983), namely $\rm k(\nu )= k(\nu
_{\circ  })\   (\nu  /\nu   _{\circ  })^{\beta}$,  where   $\rm  k(\nu
_{\circ})=10\ cm^{-2}\ gr$ and $\nu _{\circ } = 1196$ GHz (250 $\mu $m).

For a  galactocentric radius  of $\rm r=20''$,  where most of  the FIR
emission arises  (Telesco et al.   1984), we estimate $\rm  S_{\rm CO}
=(2800\pm  800)$ Jy km  s$ ^{-1}$  (from the  $ ^{12}$CO  J=1--0 channel
maps, Helfer \&  Blitz 1995), $\rm S_{\rm HI} =  (0.82\pm 0.20)$ Jy km
s$ ^{-1}$ and $\rm S_{450} = (8.74\pm 1.95)$ Jy, $\rm S_{850}=(1.23\pm
0.13)$~Jy. For $\beta =2 $ and  $\rm X_{\rm co}=5\ \rm M_{\odot }\ (K\
km  \ s^{-1}  \ pc^2)^{-1}$  (i.e. the  standard galactic  value), the
observed 850 $\mu $m flux density implies

\begin{equation}
\frac{\rm M_{\rm g}}{\rm M_{\rm d}}\approx 315 \left(e^{\rm 16.8/T}-1\right)^{-1},
\end{equation}

\noindent
For $\rm T=  25$ K  yields $\rm  M_{\rm g}/M_{\rm  d}\sim 330$,
similar to the molecular gas-to-dust ratio found in spirals using IRAS
100 $\mu $m and 60 $\mu $m fluxes (\cite{Yo86}; \cite{Yo89}).

Assuming that the  actual gas-to-dust ratio in NGC  1068 is similar to
that  for  the  Milky  Way,  namely  100-150,  we  conclude  that  the
application  of equation (3)  leads to  a significant  overestimate of
this ratio for the inner region ($\rm r\leq 20''$) of NGC 1068.  There
is indeed  independent evidence (e.g.   Downes \& Solomon  1998) which
suggests  that  the standard  galactic  conversion  factor applied  to
starburst  nuclei leads  to  significant overestimates  the $\rm  H_2$
content  in  these regions.   In  addition,  Papadopoulos \&  Seaquist
(1998) provide indications  that $\rm M(H_2)$  in the inner  region of
NGC 1068 is overestimated by a factor of two, which would then produce
agreement between  the ratio for the  nuclear region NGC  1068 and the
Milky way.

   For $r  > 20''$,  no bright  CO emission is  detected (see  Fig. 2,
Helfer  \& Blitz~1995).  There  is,  however, evidence  for faint  CO
emission. By integrating  over the area $\rm 20''\leq  r \leq 60''$ we
find  $\rm S_{\rm CO}=(1570\pm  470)$ Jy  km s$  ^{-1}$.  Over  the same
area,  we estimate $\rm  S(HI) =  (16\pm 1)$~Jy~km~s$ ^{-1}$  and $\rm
S_{450} = (4.0\pm 0.9)$ Jy, $\rm S_{850} = (0.85\pm 0.09)$ Jy. Using these
figures, together  with equation (2)  and the standard value  for $\rm
X_{\rm co}$, we obtain,

\begin{equation}
\frac{\rm M_{\rm g}}{\rm M_{\rm d}}\approx 300 \left(e^{\rm 16.8/T}-1\right)^{-1},
\end{equation}

\noindent
For $\rm  T\sim 10-15$ K we  obtain $\rm M_{\rm  g}/M_{\rm d}= 70-150$
and $\rm  M(H_2)/M(HI)\sim 5$.  The good agreement  of the gas-to-dust
ratio with  that of the Milky  way suggests that most  of the gas/dust
mass is accounted for.

However, the CO becomes exceedingly faint and essentially undetectable
at $r\geq 30''$.  If we assume the lowest gas/dust temperature of $\rm
T\sim 10$ K  allowed by the $\rm S_{450}/S_{850}$  ratio, the inferred
gas-to-dust ratio for  $30''\leq r \leq 60''$ would  be $\sim 40$ with
$\rm M(H_2)/M(HI)\sim 1$ (using the standard $\rm X_{\rm co}$).  Thus,
the  use  of  the  standard  conversion  factor  in  this  region  may
underestimate $\rm M(H_2)$ by a factor of $\sim 5$.  It seems possible
that this molecular  gas phase is not CO-bright  mainly because of low
densities  ($\rm n(H_2)<  10^3$~cm$ ^{-3}$)  rather than  low gas/dust
temperatures.   A  thermalized $  ^{12}$CO  J=1--0  line would  remain
luminous as long as $\rm T\geq \Delta \rm E_{10}/k\sim 5$ K, and lower
temperatures would  yield $\rm S_{450}/S_{850}\leq 1.13$  which is not
observed. Assuming  similar spatial  and velocity filling  factors, we
estimate that the average $ ^{12}$CO J=1--0 brightness temperature for
the  outer regions  $\rm  T_{\rm b}  \sim  0.07\times T_{\rm  b}(r\leq
20'')$.  For the various molecular  clouds (size $\sim 200$ pc) within
the $\rm  r\sim 20''$ radius  $\rm T_{\rm b}(r\leq  20'')\sim 10-20$~K
(e.g.   Planesas,  Scoville \&  Myers~1991),  hence  for similar  size
molecular  clouds in  the HI-bright  regions it  would be  $\rm T_{\rm
b}\sim  1$   K.   This  is  significantly  lower   than  the  gas/dust
temperature inferred by the  $\rm S_{450}/S_{850}$ ratio over the same
regions, thus implying sub-thermally excited $ ^{12}$CO J=1--0.

These  characteristics of  the  diffuse $\rm  H_2$  phase makes  sub-mm
  measurements very  valuable since,  while CO is  very faint  and may
  underestimate the $\rm H_2$  mass, the dust/gas temperature is still
  high enough to  allow sub-mm imaging to reveal  the distribution and
  mass  of the  $\rm H_2$  mixed with  HI, under  the assumption  of a
  canonical  gas-to-dust ratio.   We expect  this  gas phase  to be  a
  general feature  in spiral galaxies and sensitive  sub-mm imaging of
  the regions with high HI column densities is a new tool to study its
  properties through the associated dust emission.

\section{Conclusions}

The  results  of  deep   sub-mm  imaging  of  the  archetypal  Seyfert
2/starburst galaxy NGC~1068 can be summarized as follows:

1. Extended sub-mm  emission due  to dust with  a temperature  of $\rm
   T\sim  10-15$~K  is found  associated  with  the  regions with  the
   highest  HI column  densities.  A  more spatially  concentrated and
   warmer  ($\rm T\sim  20-30$  K)  component is  found  in the  inner
   starburst region where most of the gas is molecular.  The estimated
   gas-to-dust  ratio for the  inner region  is $\rm  M_{\rm g}/M_{\rm
   d}\sim 330$ while for the  dust emission associated with the HI gas
   we find $\rm M_{\rm g}/M_{\rm d}\sim 70-150$ depending on the value
   of the gas/dust temperature.

2.  Under  the  assumption  that  a  Milky Way  value  of  $\rm  M_{\rm
    g}/M_{\rm d}\sim  100-150$ applies also  to NGC 1068,  we conclude
    that the high value of  $\rm M_{\rm g}/M_{\rm d}$ in its starburst
    region is a  result of an overestimate of  the molecular gas mass.
    This  seems  to be  a  systematic  effect  of using  the  standard
    galactic conversion factor to convert $ ^{12}$CO J=1--0 luminosity
    to molecular gas mass in starburst environments.

3. The  diffuse $\rm  H_{2}$ gas  that  is mixed  with the  HI gas  is
   associated with  sub-thermally excited and very  faint CO emission.
   Furthermore it is more  massive than HI, with $\rm M(H_2)/M(HI)\sim
   5$ and  the standard  galactic conversion factor  may underestimate
   $\rm M(H_2)$ at low gas/dust temperatures of $\rm T\sim 10$ K.  For
   this gas phase sensitive sub-mm imaging provides a better tool than
   CO imaging in revealing its distribution and mass.

\acknowledgements

We thank Rob Ivison and Lorne  Avery for conducting a wonderful set of
observations  with SCUBA,  and Wayne  Holland for  friendly  and useful
advice.  We are grateful to Elias  Brinks for the HI map and Tamara Helfer
for the $ ^{12}$CO J=1--0 map.   E.  R.  S acknowledges the support of
a research  grant from the  Natural Sciences and  Engineering Research
Council of~Canada.

\newpage

\figcaption{\underline{Grey scale}: Velocity-integrated HI brightness
with a range  of $100-830$ mJy beam$ ^{-1}$ km  s$ ^{-1}$, and $\sigma
_{\rm   rms}\sim   100$  mJy   beam$   ^{-1}$   km   s$  ^{-1}$.    \\
\underline{Contours}:  850  $\mu  $m  brightness (top),  450  $\mu  $m
brightness (bottom).  The contours are: \\  (1, 2, 3, 4, 5, 7, 10, 15,
20, 25, 30, 35)$\times \sigma$,  where $\sigma _{\rm rms}(850\ \mu \rm
m)=10$  mJy   beam$  ^{-1}$  and  $\sigma  _{\rm   rms}(450\  \mu  \rm
m)=90$~mJy~beam$  ^{-1}$.  The  resolution of  all the  maps  is $\sim
15''$.
\label{fig1}}

\figcaption{\underline{Gray  scale}:  Velocity-integrated  $  ^{12}$CO
J=1--0 brightness from Helfer \& Blitz 1995, convolved to a resolution
of  $\sim  9''$, and  a  range  of $100-307$  Jy  beam$  ^{-1}$ km  s$
^{-1}$.\\ \underline{Contours}:  The bright 450 $\mu $m  emission at a
resolution of  $\sim 9''$, the contours  are: (8, 10, 12,  14, 16, 18,
20, 22, 24, 26, 28, 30)$\times  \sigma$, where $\sigma = 50$ mJy beam$
^{-1}$.
\label{fig2}}

\newpage
\centerline{\large Table 1}
\centerline{\large Calibration data}
\begin{center}
\begin{tabular}{ c c c c c c} \hline\hline

Wavelength & $\theta _{\rm HPBW}$  & $\rm S_{\rm tot}(cal) ^{\rm a}$ & 
G, $\delta \rm G$/G $ ^{\rm b}$ & $\sigma _{\rm rms} ^{\rm c}$ & ($\rm f_{20}$,\
 \ \ $\rm f_{60}$) $ ^{\rm d}$\\ \hline\hline

850 $\mu $m & $15.25''$ & $4.56\pm 0.17$  & 265, 15\% & 10 & (1.03, 1.30) \\
450 $\mu $m  & $8.75''$  & $11.5\pm 1.2$  & 670, 30\% & 50 & (1.45, 2.45) \\ \hline

\end{tabular}
\end{center}

$ ^{\rm a}$ The total flux of CRL 618 in Jy, taken from the JCMT secondary 
calibrators list, \hspace*{1.0cm} Sandell 1998.

$ ^{\rm b}$ The flux gain in Jy/beam Volts$ ^{-1}$ and its fractional 
uncertainty estimated from an \hspace*{1.0cm} extensive set of beam maps 
and photometry on CRL 618 and Mars.

$ ^{\rm c}$ The thermal rms error of the final maps in mJy/beam.

$ ^{\rm d}$ The flux correction factor for circular area with $r=20''$
 and  $r=60''$ radius,  estimated  \hspace*{1.0cm} from  beam maps  of
 CRL~618  (see Sandell  1998).  If  S is  the integrated  flux density
 \hspace*{1.0cm} within  that area,  the error-beam corrected  flux is
 $\rm S_{\rm c} = S/f$.

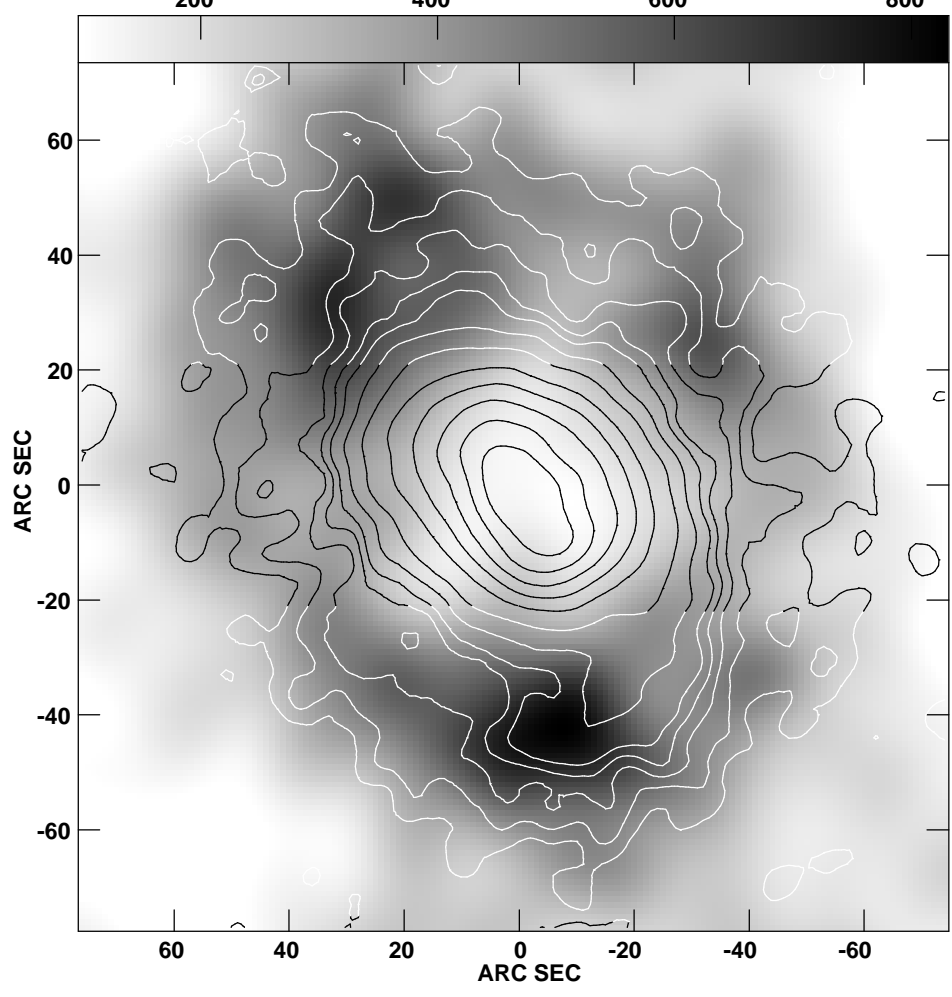
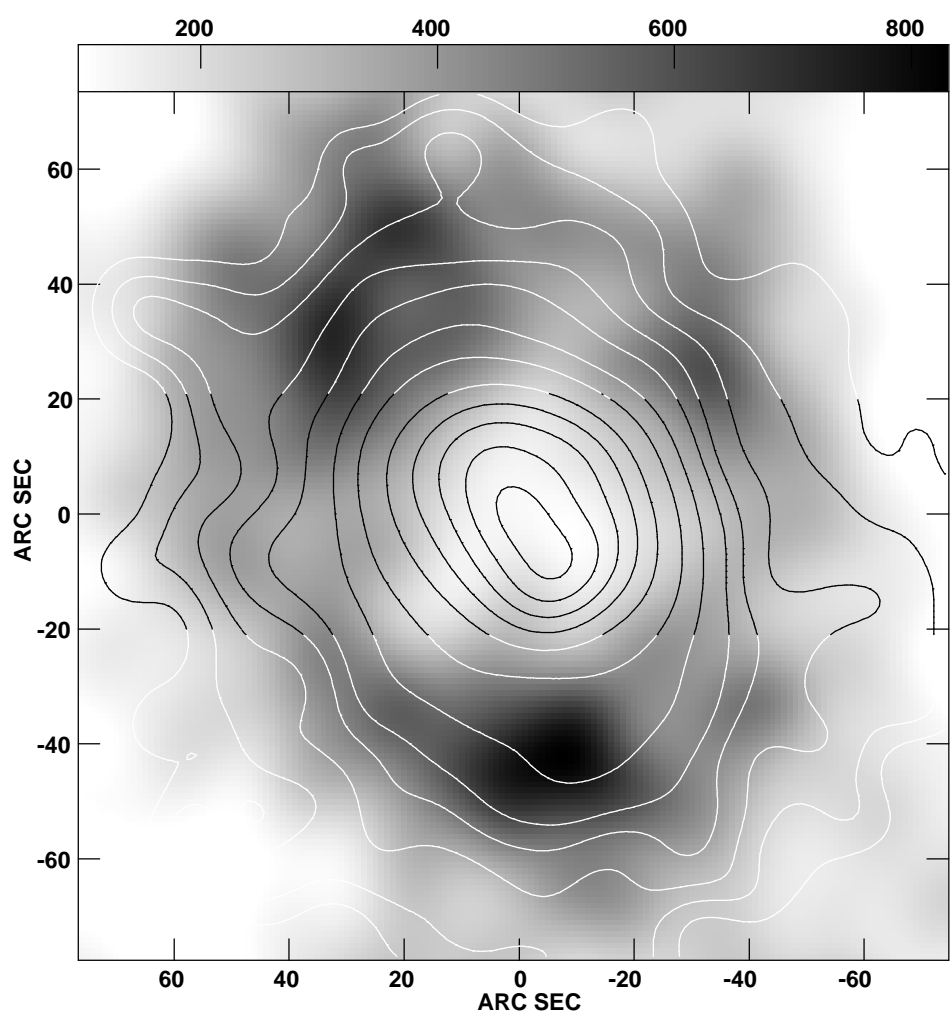
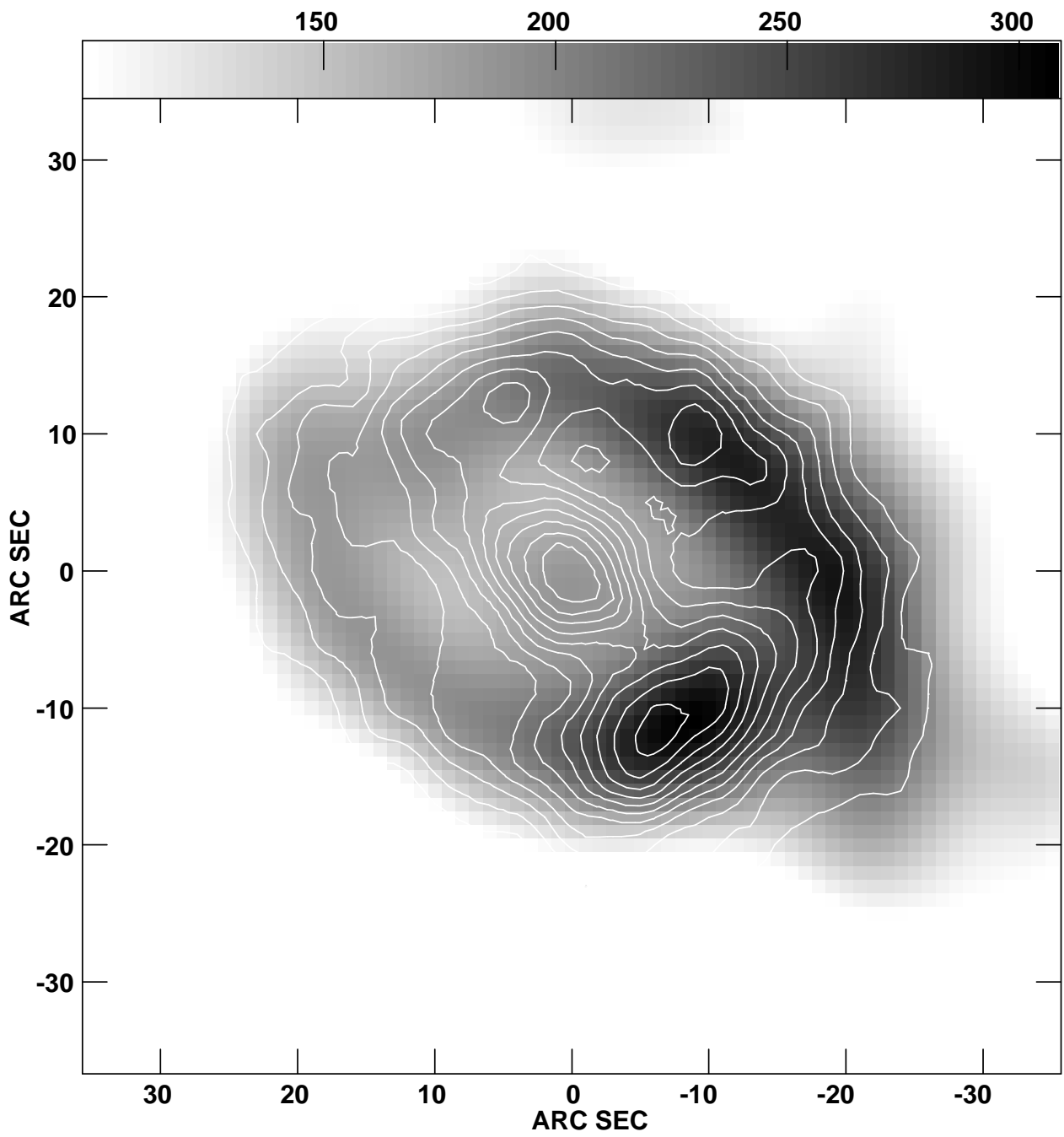
\end{document}